\documentclass[10pt]{elsarticle}
\usepackage{graphics, graphicx, amsmath, amssymb}
\usepackage[usenames,dvipsnames]{color}
\usepackage{verbatim}
\usepackage{lineno,hyperref}
\definecolor{Red}{rgb}{1,0,0}

{\catcode`\|=\active
 \gdef\Braket#1{\begingroup
\mathcode`\|32768\let|\BraVert\left<{#1}\right>\endgroup}}
\def\BraVert{\egroup\,\mid\,\bgroup}

\begin{document}

\begin{frontmatter}

\title{Noise Sensing via Stochastic Quantum Zeno}

\author{Matthias M. M\"uller\fnref{myfootnote}}
\address{Peter Grünberg Institute - Quantum Control (PGI-8), Forschungszentrum J\"ulich GmbH, D-52425 Germany}
\fntext[myfootnote]{ma.mueller@fz-juelich.de}

\author{Stefano Gherardini}
\author{Nicola Dalla Pozza}
\author{Filippo Caruso}
\address{Dept.\,of Physics and Astronomy \& European Laboratory for Non-Linear Spectroscopy (LENS), University of Florence, Florence, Italy}

\begin{abstract}
The dynamics of any quantum system is unavoidably influenced by the external environment. Thus, the observation of a quantum system (probe) can allow the measure of the environmental features.
Here, to spectrally resolve a noise field coupled to the quantum probe, we employ dissipative manipulations of the probe, leading to so-called Stochastic Quantum Zeno (SQZ) phenomena.
A quantum system coupled to a stochastic noise field and subject to a sequence of protective Zeno measurements slowly decays from its initial state with a survival probability that depends both on the measurement frequency and the noise. We present a robust sensing method to reconstruct the unknown noise power spectral density by evaluating the survival probability that we obtain when we additionally apply a set of coherent control pulses to the probe. The joint effect of coherent control, protective measurements and noise field on the decay provides us the desired information on the noise field.
\end{abstract}

\begin{keyword}
Open quantum systems, quantum noise sensing, stochastic quantum Zeno effect
\end{keyword}

\end{frontmatter}

\newpage

\section{Introduction}

The aim of quantum noise spectroscopy is to indirectly infer statistical features of noise fluctuation fields acting on a quantum system used as probe~\cite{Degen2017,Szankowski2017}.
Dynamical decoupling sequences~\cite{Viola_PRL_1999,KhodjastehPRL2005,UhrigPRL2007,UhrigNJP2008}, that can be seen as an extension of the Ramsey interference measurement, are the standard tool for quantum noise spectroscopy~\cite{Yuge2011,Biercuk2011,Alvarez2011,Bylander2011}.
To enhance the information content, a set of coherent control pulses has to be independently applied to the probe, such that properly designed reconstruction algorithms allow for a high reconstruction fidelity of the noise power spectral density~\cite{KofmanPRL2001,GordonJPB2007,Alvarez2011,Paz-Silva2014,Zwick2016,Mueller2018,Szankowski2018,Krzywda2019}, also including for non-Gaussian noise~\cite{Szankowski2017,Norris2016,Ramon2019,Sung2019}.
However, also quantum sensing procedures that are an extension of the Rabi measurement can be an effective approach \cite{Mueller2016b, Do2019,Sakuldee2019}.
These protocols rely on the SQZ phenomena and the probe, while it is affected by the noise field to be inferred, is subjected to a sequence of (not necessarily projective) quantum measurements, which confine the probe system in its initial state (or subspace). For a sufficiently small amount of noise and sufficiently frequent measurements, such a procedure tends to freeze the dynamics of the system~\cite{Misra1977,ItanoPRA1990,KofmanNAT2000,FisherPRL2001,FacchiPRL2002,Facchi2008,SmerziPRL2012,SchaferZeno,Signoles2014}, while for a low repetition rate, the measurements can even enhance the decay (Anti-Zeno regime)\,\cite{Goan,Hanggi,Chaudhry}.
Conversely, if the evolution of the probe is stochastic due to the presence of an external fluctuating field, and the measurements are applied at an intermediate repetition rate, the confinement of its wave-function in the initial state (subspace) slowly decays and the decay depends on the statistics of the noise field~\cite{Gherardini2016,Mueller2017,Gherardini2017}. This has opened the way to the sensing of the fluctuations of noise~\cite{Mueller2016b,Do2019,Mueller2016}. Here, we first present the general idea of the SQZ approach to quantum noise sensing and then discuss the adopted approximations and the error propagation in the spectral reconstruction.

\section{Stochastic Quantum Zeno}\label{sec:SQZD_general}

Let us consider a quantum system governed by the Hamiltonian
$H(t)=H_0 + \Omega(t) H_{n}$, where $H_0$ is the static Hamiltonian of the system and $\Omega(t)$ is a stochastic noise field coupled to the system via the operator $H_n$.

SQZ phenomena can be observed by applying a sequence of (not necessarily projective) measurements that confine the system dynamics in a subspace. In our case, we apply a sequence of $N$ projective measurements of the population in the initial state $|\psi_0\rangle$ at times $t_j$ ($j=0,\dots,N$). More details on the measurement scheme and an experimental realization can be found in~\cite{Do2019,Gherardini2017}. The key feature is to throw away the measurement outcome if we do not find the system in $|\psi_0\rangle$ and keep only the outcome associated to $|\psi_0\rangle$. This effectively resets the state of the system to $|\psi_0\rangle$, while the population in $|\psi_0\rangle$ reduces at each step (i.e., after each measurement) by the factor
\begin{equation}
 q_j = \left|\left\langle \psi_0\left|\mathcal{T}\exp\left(-i\int_{t_{j-1}}^{t_j} H(t) dt\right)\right|\psi_0\right\rangle\right|^2,
\end{equation}
where $\mathcal{T}$ denotes the time-ordering operator and the reduced Planck constant $\hbar$ has been set to $1$. For small time intervals $t_j-t_{j-1}$ we can make the following second-order Dyson-series approximation:
\begin{eqnarray}
  q_j &\approx& \left|\left\langle \psi_0\left| 1- i\int_{t_{j-1}}^{t_j} H(t) dt - \int_{t_{j-1}}^{t_j} \int_{t_{j-1}}^t H(t)H(t') dt dt'  \right|\psi_0\right\rangle\right|^2
 . 
\end{eqnarray}

If we set $|\psi_0\rangle = |0\rangle$, $H_0=\Delta\,\sigma_z$ and $H_n=\sigma_x$, with the Pauli-operators $\sigma_z = |0\rangle\!\langle 0| - |1\rangle\!\langle 1|$, $\sigma_x = |0\rangle\!\langle 1| + |1\rangle\!\langle 0|$ and level splitting $\Delta$, we obtain
\begin{equation}
  H(t)=\Delta\,\sigma_z + \Omega(t)\sigma_x
\qquad\text{and}\qquad
  q_j \approx 1-\left(\int_{t_{j-1}}^{t_j}\Omega(t) dt\right)^2.
\end{equation}
It is worth noting that, due to the choice of the initial state and the measurement operator, the contribution of the level splitting $\Delta$ vanishes in the second order approximation and only the term $\Omega (t)\sigma_x$ contributes to $q_j$.

\section{Quantum Probes based on the Stochastic Quantum Zeno}
Now, let us investigate how such a two-level system can be used as a quantum probe within the framework of SQZ effect. For this purpose, we introduce an additional control field (on the $\sigma_x$ operator) so that the Hamiltonian of the two-level system with basis states $|0\rangle$ and $|1\rangle$ reads:
\begin{equation}
 H(t)=\Delta\sigma_z +\Omega_c(t)\sigma_x + \Omega_n(t)\sigma_x
\end{equation}
where $\Omega_c(t)$ and $\Omega_n(t)$ denote, respectively, the control and stochastic noise fields. The initial state of the system is again $|0\rangle$.
Therefore, according to the second-order Dyson-series approximation in section \ref{sec:SQZD_general}, one obtains
\begin{eqnarray}
 q_j=1-\left(\int_{t_{j-1}}^{t_j}\Omega_c(t)+\Omega_n(t)dt\right)^2.
\end{eqnarray}

\subsection{Survival Probability: Contribution of Noise and Control}\label{sec:SurvivalProbability}
While in the previous subsections we have focused on the survival probability after one single measurement, here, we study the overall survival probability after applying a sequence of $N$ measurements, that is simply given by the product of the single measurement survival probabilities $q_j$\,:
\begin{eqnarray}
 P=\prod_{j=1}^N q_j
 \approx \exp\left[  -\sum_{j=1}^N \left(\int_{t_{j-1}}^{t_j}\Omega_c(t)+\Omega_n(t)dt\right)^2  \right].
\end{eqnarray}
Hence, $P$ factorizes in three contributions, i.e.
\begin{equation}
 P=P_{\rm n}P_{\rm c}P_{\rm cn}\,,
\end{equation}
where $P_{n}$ only depends on the noise, $P_{c}$ only on the control and $P_{cn}$ on both. They are given by the following expressions:
\begin{eqnarray}
&\displaystyle{P_{n}=\exp\left[ -\sum_{j=1}^N \left(\int_{t_{j-1}}^{t_j}\Omega_n(t)dt\right)^2 \right],\,\,\,\,\,\,
P_{c}=\exp\left[ -\sum_{j=1}^N \left(\int_{t_{j-1}}^{t_j}\Omega_c(t)dt\right)^2 \right]},&\nonumber \\
&\displaystyle{P_{cn}=\exp\left[-2\sum_{j=1}^N \left(\int_{t_{j-1}}^{t_j} \Omega_c(t)dt \right) \left(\int_{t_{j-1}}^{t_j} \Omega_n(t')dt' \right) \right].}&
\end{eqnarray}
If the noise is \textit{weak} enough, i.e., $\displaystyle{\sum_{j=1}^N \left(\int_{t_{j-1}}^{t_j}\Omega_n(t)dt\right)^2\ll 1}$, then we can neglect the factor $P_{n}\approx 1$. The term $P_{c}$, instead, depends only on the control pulse and, thus, it can be directly calculated. Finally, the term $P_{cn}$ is a cross-term of noise and control and contains all the interesting information that we can extract on the fluctuating noise field.

Now, we have to understand in more detail, what exactly we can learn about the noise from the value of $P_{cn}$. We start by defining
\begin{eqnarray}
 \tilde\Omega_c (t) \equiv \sum_{j=1}^N \left(\int_{t_{j-1}}^{t_j} \Omega_c(t')dt'\right) w_j(t)\,,\,\text{ with}\qquad w_j(t)= \begin{cases}
  1\quad t_{j-1}\leq t <t_j\\
  0\quad\text{otherwise.}                                                                                                                                                                                                                                 \end{cases}
\end{eqnarray}
By this definition $\tilde\Omega_c (t)$ is the piece-wise average of $\Omega_c(t)$ in each interval between two measurements, and we can write
\begin{eqnarray*}
 \sum_{j=1}^N \left(\int_{t_{j-1}}^{t_j} \Omega_c(t)dt \right) \left(\int_{t_{j-1}}^{t_j} \Omega_n(t')dt' \right)   
 =\sum_{j=1}^N  \left(\int_{t_{j-1}}^{t_j} \left(\int_{t_{j-1}}^{t_j} \Omega_c(t)dt \right) \Omega_n(t')dt' \right)   \\
 =\sum_{j=1}^N  \left(\int_{t_{j-1}}^{t_j} \tilde\Omega_c(t') \Omega_n(t')dt' \right)   
 = \int_{t_0}^{t_N} \tilde\Omega_c(t) \Omega_n(t)dt\,.
 \end{eqnarray*}
In this way, $P_{\rm cn}$ can be simply expressed in the following more compact form:
\begin{equation}
P_{\rm cn} = \exp\left[-2\int_{t_0}^{t_N} \tilde\Omega_{c}(t)\Omega_{\rm n}(t)dt\right].
\end{equation}

\section{Estimation of the Noise Correlation Function}\label{sec:measurement}

As we have seen in section \ref{sec:SurvivalProbability}, a population measurement at the end of the Zeno sequence of the remaining population in state $|0\rangle$ will provide the survival probability $P=P_{n}P_{c}P_{cn}$. As discussed above, $P_{cn}$ is an unknown quantity depending on the correlation between noise and control and contains the information on the noise field to be inferred. Indeed, one has that $P_{cn} \approx P/P_{c}$, and
\begin{equation}\label{log_P_over_Pc}
 \int_{t_0}^{t_N} \tilde\Omega_c(t)\Omega_n(t)dt \approx -\frac{1}{2}\ln\frac{P}{P_{\rm c}}\,.
\end{equation}

We can summarize the noise contribution in the noise autocorrelation function
$g(t-t') \equiv \left\langle\Omega_n(t)\Omega_n(t')\right\rangle$. Thereby, we assume that the noise field is described by a stationary stochastic process, such that the autocorrelation function $g(t-t')=g(\tau)$ depends only on the time difference $\tau \equiv t' - t$. 
To obtain this correlation function from the left side of Eq.~\eqref{log_P_over_Pc}, we have to square and average, so that  
\begin{eqnarray}
\left\langle \left(\int_{t_0}^{t_N} \tilde\Omega_c(t) \Omega_n(t)dt\right)^2 \right\rangle 
= \int_{t_0}^{t_N}\int_{t_0}^{t_N} \tilde\Omega_c(t)\tilde\Omega_c(t') \left\langle\Omega_n(t)\Omega_n(t')\right\rangle dt dt' \equiv \chi_{N}^{(2)}\,,\nonumber\\
\end{eqnarray}
where we have defined the function
$\chi_N^{(2)}$ that quantifies the correlation of noise and control. We call it \textit{second-order decoherence function}. Since the control field $\tilde \Omega_c(t)$ is not stochastic, the average is performed only over the noise contribution.

If we now want to relate $\chi_N^{(2)}$ to the measurement of the survival probability $P$, we have to distinguish two cases that depend on the physical implementation of the probe. Indeed, while in many cases quantum measurements of a two-level system yield just the outcomes $0$ or $1$ according to the probability $P$, in some cases also a single measurement can yield the real numbered value of $P$ with high precision. Let us first turn to the latter case:

For instance, in Ref.~\cite{Do2019} the quantum probe was realized by hyperfine levels of a BEC on an atom-chip. While the Zeno projective measurements were realized by laser pulses, the final measurement of $P$ is performed by a Stern-Gerlach experiment that separates the atoms according to the electronic state they are projected in. Such a measurement can give the value of $P$ up to a few percent and the error depends on technical imperfections and the shot noise related to the number of atoms in the BEC. A similar measurement can be envisioned in other collective sensors, e.g. through fluorescence in ensembles of NV-centers~\cite{Balasubramanian2019,Zhou2019}.

If we assume such a measurement of $P$, each time we repeat the Zeno sequence, we obtain a value for $\frac{1}{2}\ln\frac{P}{P_{c}}$.
As recently shown in~\cite{Do2019}, by averaging the square of the integral in Eq.\,(\ref{log_P_over_Pc}) with respect to the noise realizations, one gets
\begin{eqnarray}\label{eq:decoherence_function}
\frac{1}{4}\left\langle \ln^2 \frac{P}{P_{\rm c}} \right\rangle & \approx & \frac{1}{4}\left\langle \ln^2 P_{\rm cn} \right\rangle =\chi_N^{(2)}\,.
\end{eqnarray}

Conversely, if we assume that a measurement of $P$ will yield $0$ and $1$, we have access only to the averaged value $\langle P\rangle$, where the average is performed over the measurement statistics and the realizations of the noise, so that
\begin{eqnarray}
 \left \langle \frac{P}{P_{\rm c}}\right\rangle = \langle P_{\rm cn}\rangle = \left \langle \exp\left[-2\int_{t_0}^{t_N} \tilde\Omega_{c}(t)\Omega_{\rm n}(t)dt\right] \right\rangle
\end{eqnarray}
By assuming $\int_{t_0}^{t_N} \tilde\Omega_{c}(t)\langle\Omega_{\rm n}(t)\rangle dt = 0$ we can make the second-order (in the noise contribution $\Omega_n(t)$) approximation
\begin{eqnarray}
 \left \langle \frac{P}{P_{\rm c}}\right\rangle = \langle P_{\rm cn}\rangle = \left \langle \exp\left[-2\int_{t_0}^{t_N} \tilde\Omega_{c}(t)\Omega_{\rm n}(t)dt\right] \right\rangle\\
 =\left\langle 1- 2\int_{t_0}^{t_N} \tilde\Omega_{c}(t)\Omega_{\rm n}(t)dt + \frac{4}{2} \int_{t_0}^{t_N}\int_{t_0}^{t_N} \tilde\Omega_c(t)\tilde\Omega_c(t') \Omega_n(t)\Omega_n(t') dt dt' + ... \right\rangle\\
 \approx \exp\left[ 2 \int_{t_0}^{t_N}\int_{t_0}^{t_N} \tilde\Omega_c(t)\tilde\Omega_c(t') \left\langle\Omega_n(t)\Omega_n(t')\right\rangle dt dt'   \right] = \exp\left[ 2\chi_N^{(2)}\right] \label{eq:average_P} 
\end{eqnarray}
A similar approximation was found also in~\cite{Mueller2017} for the survival probability of a many-body system within a subspace subject to stochastic Zeno protection measurements in the weak Zeno regime.
Moreover, it can be shown that this holds also for Gaussian noise since there the contribution of the higher moments (or more precisely cumulants) to the decoherence function vanish~\cite{Szankowski2017}.
Note that Eq.~\eqref{eq:average_P} is slightly different from Eq.~\ref{eq:decoherence_function} and whenever we refer to the specific way of obtaining $\chi_N^{(2)}$ from the measurement of $P$ in the remainder of the article we will refer to Eq.~\ref{eq:decoherence_function}.

Now, let us introduce the noise power spectral density $S(\omega)$ through the inverse Fourier transform
  $\displaystyle{g(\tau)=\frac{1}{2\pi}\int_{0}^{\infty} S(\omega) \mathrm{e}^{i\omega\tau}d\omega}$,
with the result that the second-order decoherence function in the spectral form equals to
\begin{equation}
 \chi_N^{(2)} = \frac{1}{2\pi}\int_{0}^{\infty}\int_0^{t_N}\int_0^{t_N} S(\omega)\mathrm{e}^{i\omega (t-t')}\tilde\Omega_c(t)\tilde\Omega_c(t') dt dt'd\omega\,.
\end{equation}
Then, if we introduce also the Fourier transform of the control field, i.e.,
 $Y(\omega) \equiv \int_0^{t_N} \tilde\Omega_c(t')e^{i\omega t'}dt'$,
and the filter function $F(\omega) \equiv \frac{1}{2\pi}|Y(\omega)|^2$, the second-order decoherence function becomes
\begin{equation}
 \chi_N^{(2)} = \int_{0}^\infty S(\omega)F(\omega)d\omega\,,
\end{equation}
similarly to the Ramsey-type dynamical decoupling case\,\cite{Degen2017,Szankowski2017,KofmanPRL2001}.
In this way, by measuring $\chi_N^{(2)}$ for different choices of the control field $\tilde\Omega_c(t)$, we can obtain the noise power spectral density $S(\omega)$ in different frequency regimes and, thus, reconstruct its functional shape\,\cite{Alvarez2011,Mueller2018,Krzywda2019,Do2019} as desired.

\subsection{Higher order correlation functions}

We can also generalize the approach presented above (for the case where we can measure $P$ in a single-shot read-out like in the experiment of \cite{Do2019}) to correlation functions of arbitrary order. To this end, we introduce the $k$-th order decoherence function
\begin{eqnarray}
 &\chi_{N}^{(k)} = \frac{1}{4}\left\langle \ln^k P_{cn} \right\rangle&\nonumber \\
 &=\displaystyle{\int_{t_0}^{t_N}\dots\int_{t_0}^{t_N} \tilde\Omega_c(t^{(1)})\dots\tilde\Omega_c(t^{(k)})
 \left\langle\Omega_n(t^{(1)}) \cdots \Omega_n(t^{(k)}) \right\rangle dt^{(1)} \cdots dt^{(k)}}&
\end{eqnarray}
that allows to evaluate the $k$-point noise correlation function
\begin{equation}
g^{(k)}(t^{(1)},\dots,t^{(k)}) \equiv \left\langle\Omega_n(t^{(1)}) \dots \Omega_n(t^{(k)}) \right\rangle \,.
\end{equation}
Note that the measurement of higher order noise correlation functions can be very important in cases where the noise statistics is not characterized completely by the second-order correlation, e.g., for non-Gaussian noise fields.

\section{Statistical and experimental error analysis}
Since $P_{cn}$ is a stochastic variable and takes a different value each time the sensing procedure is repeated, we can discuss it in terms of ergodicity. The main reasons for the stochasticity are:\\
(i)~finite duration of the sensing protocol and stochasticity of the noise field,\\
(ii)~imperfections in the probe dynamics and control pulses and the application of non-ideal intermediate projective measurements (implementation of $F(\omega)$),\\
(iii)~errors in the measurements of $P$.

In our sensing protocol, the original spectral density $S^{(orig)}$ has to be reconstructed from the measurements of $P$, leading to its reconstruction $S^{(rec)}$. We start by choosing $K$ different filter functions $F_k(\omega)$ ($k=1,\dots, K$). For each filter function we perform $M$ times the time dynamics and final measurement of $P$. Ideally, each time we would measure
\begin{eqnarray}
\chi_k=\int_0^\infty S^{(orig)}(\omega) F_k(\omega) d\omega .
\end{eqnarray}
Due to the aforementioned errors, instead, we measure $\chi_{k,m} = \chi_k + \Delta \chi_{k,m}$, where $\Delta\chi_{k,m}$ ($m=1,\dots, M$) encodes the cummulative error in the measurement. We can calculate the mean and standard deviation of $\chi_{k,m}$ as
\begin{eqnarray}
 \overline{\chi}_k\equiv\frac{1}{M}\sum_{m=1}^M \chi_{k,m}\quad\text{and}\quad \Delta\chi_k \equiv \sqrt{\frac{1}{M-1}\sum_{m=1}^M (\chi_{k,m} - \overline{\chi}_k)^2}\,.
\end{eqnarray}
To reconstruct $S^{(rec)}(\omega)$ from the measurements, we have to calculate the Gramian matrix $A_{kl}=\int_0^\infty F_k(\omega) F_l(\omega) d\omega$~\cite{Mueller2018,DallaPozza2019}. This symmetric matrix can be orthogonalized as $A= V^T \Lambda V$, where $V$ contains the eigenvectors of $A$ and $\Lambda=\mathrm{diag} (\lambda_1,\dots, \lambda_K)$ the eigenvalues. Then, the noise power spectral density can be reconstructed as
\begin{eqnarray}
 S^{(rec)}(\omega) = \sum_{k=1}^K \overline{\chi}_k \tilde{F}_k(\omega),
\end{eqnarray}
where the $\tilde{F}_k(\omega) \equiv \sum_{i,l=1}^K \frac{1}{\lambda_l}V_{lk}V_{li}F_i(\omega)$ are the transformed filter functions (i.e., corrected by the inverse Gramian matrix). As a consequence, the reconstruction error can be estimated as
\begin{eqnarray}\label{eq:reconstruction_error}
 \big|S^{(rec)}(\omega) - S^{(orig)}(\omega)\big| \approx  \Delta S^{(rec)}(\omega) \equiv \sum_{k=1}^K \tilde{F}_k(\omega)\Delta {\chi}_k\,.
\end{eqnarray}
Now, we can analyze the precision of the reconstruction given by Eq.~(\ref{eq:reconstruction_error}). First of all, we obviously obtain an error from the reconstruction of the spectral density in a truncated space. This depends on the choice of filter functions (spectral shape and number) and the shape and bandwidth of the noise power spectral density. However, within the subspace spanned by the filter functions, the reconstruction error exclusively arises from the $\Delta \chi_{k,m}$. Here, we have already named three main contributions in the beginning of this section. 

\begin{figure}[t]
 \centering
 \includegraphics[width=0.9\textwidth]{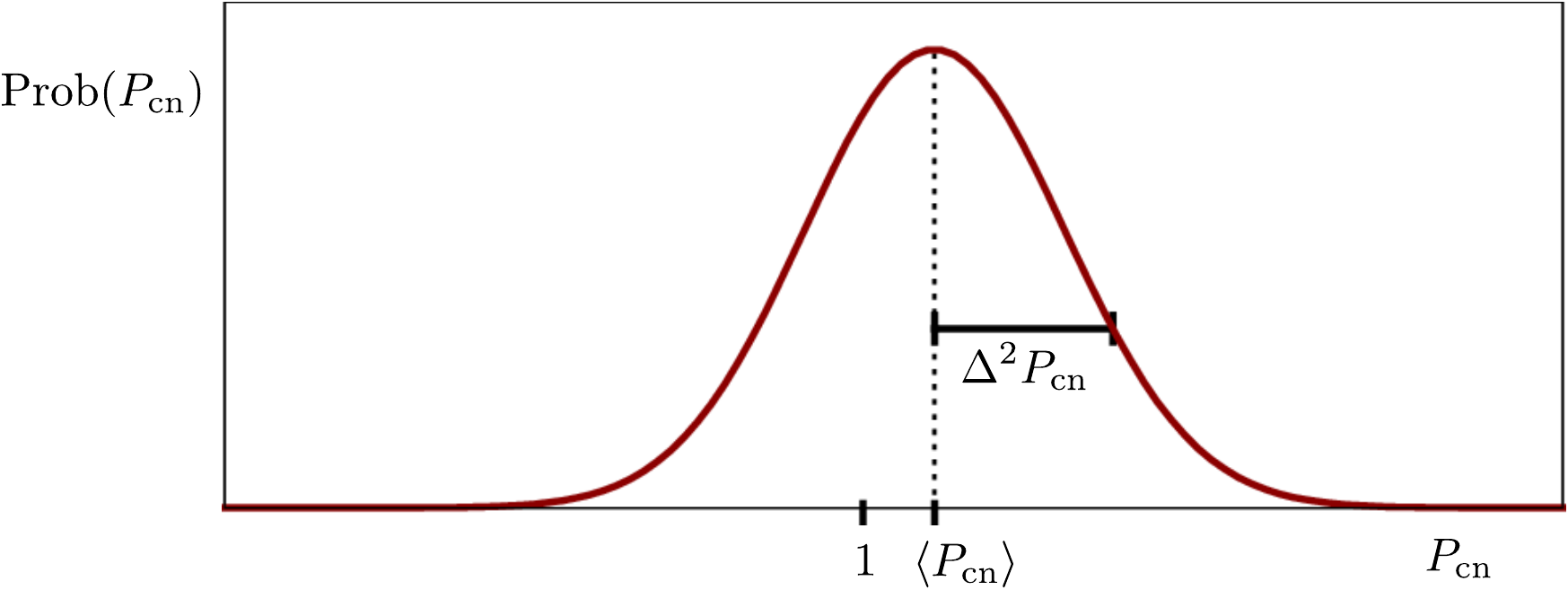}
 \caption{Mean and variance detection of the noise contribution. The figure shows the probability distribution of $P_{\rm cn}$ that contains all available information on the noise correlation function. The main effects are an offset of the average value $P_{\rm cn}$ from $1$ and the finite variance of the distribution. Depending on the sensor one can either access both effects through Eq.~\eqref{eq:decoherence_function} or just the offset value $\langle P_{\rm cn}\rangle$ when evaluating $\chi_N^{(2)}$ through Eq.~\ref{eq:average_P}.}
 \label{fig:ergodicity}
\end{figure}

If we ignore for a moment errors (ii) and (iii) and focus on (i), we can make the following considerations concerning the possible \textit{breaking of the ergodic hypothesis} of the system-environment interaction modes~\cite{Gherardini2017}. In this regard, if the duration $t_{N}$ of the single experiment is long enough to effectively reduce the stochasticity due to error (i), then one could just find, also experimentally, that
\begin{equation}\label{ergodic_hyp}
P_{\rm cn} \approx \langle P_{\rm cn}\rangle\,.
\end{equation}
The validity of Eq.\,(\ref{ergodic_hyp}) corresponds to an ergodic hypothesis in the following sense: We can consider $P_{\rm cn}$ as a time average over the noise since it essentially depends on the integral $\int_0^{t_N}\tilde\Omega_{c}(t)\Omega_{\rm n}(t)dt$. On the other hand, $\langle P_{\rm cn}\rangle$ is the average over many realizations of the noise. If for a long enough sensing time $t_N$ Eq.~\eqref{ergodic_hyp} holds, we have an equality between the ensemble and time average of $P_{\rm cn}$ and furthermore $\Delta^2 P_{\rm cn} \equiv \langle P_{\rm cn}^2\rangle - \langle P_{\rm cn}\rangle^2 = 0$, i.e. $P_{\rm cn}$ does not depend on the single realization anymore. The ensemble average is ideally computed over an infinite number of protocol realizations with finite duration $t_{N}$, while the time averaged is derived by applying once an ideally long sensing procedure. If Eq.~\eqref{ergodic_hyp} does not hold, instead we find $\Delta^2 P_{\rm cn} > 0$ and before each measurement $P_{\rm cn}$ takes a value according to some probability distribution Prob$(P_{\rm cn})$. Fig.~\ref{fig:ergodicity} shows a pictorial representation of this distribution. Now, we can gain some insight into the difference between the two measurement approaches discussed in section~\ref{sec:measurement} by assuming $|1-P_{\rm cn}|\ll 1$.
We find from Eq.\,\eqref{eq:decoherence_function} that
\begin{eqnarray}
\chi_N^{(2)} = \frac{1}{4} \left\langle \ln^2 P_{\rm cn} \right\rangle \approx \frac{1}{4}\left(\Delta^2 P_{\rm cn} + (\langle P_{\rm cn}\rangle -1)^2\right),
\end{eqnarray}
while from Eq.\,\eqref{eq:average_P} we find that
\begin{eqnarray}
 \chi_N^{(2)}=\frac{1}{2}\ln \langle P_{\rm cn}\rangle \approx \frac{1-\langle P_{\rm cn}\rangle}{2}
\end{eqnarray}
for the two different measurement approaches. Furthermore, the ensemble average and variance under this assumption can be expressed as
\begin{eqnarray}
 \Delta^2 P_{\rm cn} \approx 4 \chi_N^{(2)} \quad\text{and}\quad  \langle P_{\rm cn}\rangle-1\approx 2\chi_N^{(2)}
\end{eqnarray}
to first order in $\chi_N^{(2)}$. This leads us to the fact that the measurement of the decoherence function $\chi_N^{(2)}$ in the approach of Eq.~\eqref{eq:decoherence_function} is based on both the mean value and the variance of $P_{\rm cn}$ since we can evaluate $P_{\rm cn}$ in every single measurement. Conversely, when we choose the approach of Eq.~\eqref{eq:average_P}, where we assume only outcomes 0 and 1, we have access only to the average value of $P_{\rm cn}$ and not the information hidden in the width of Prob$(P_{\rm cn})$. Furthermore, at least in the regime of $|1-P_{\rm cn}|\ll 1$, the quantities $\Delta^2 P_{\rm cn}$ and $\langle P_{\rm cn}\rangle$ are not independent and we can measure the noise spectrum only due to the presence of a variance between the single realizations. In a similar setting, this finding was also called a \textit{breaking of the ergodic hypothesis}\cite{Gherardini2017}.

If we go back to the errors (ii) and (iii), stemming from the imperfect realization of the sensing protocol and the finite precision of the final quantum measurement, respectively, are purely technical errors. With the advance of quantum technological platforms, these contributions to the error will continuously decrease and thus, the main source of error in Eq.~(\ref{eq:reconstruction_error}) will decrease as well. In a recent experimental implementation of this protocol~\cite{Do2019} with a BEC\,\cite{Ketterle} on a chip, it has been shown that already present technology can lead to a small reconstruction error $||\Delta S(\omega)||/||S^{(orig)}(\omega)||<0.1$ (where $||\cdot||$ is the $L_2$-norm) and that this value could be further decreased by more than an order of magnitude by improving the precision of implementation and measurement.

\section{Discussion and conclusions}
We have analyzed in depth a recently introduced quantum sensing technique based on the SQZ effect. In particular, with respect to Ref.\,\cite{Do2019}, we have provided new mathematical details, especially regarding the formal expression of the survival probability $P$ and the measure of high-order noise correlation functions. Moreover, we have also discussed how to possibly characterize at the experimental level the influence of non-modeled sources of errors in relation with ergodicity breaking conditions and technical imperfections. As outlook, these results represent further steps towards other Zeno-based noise sensing schemes where, for instance, the noise is non-Gaussian or non-Markovian probe dynamics play a role.

\vspace{0.4cm}

\section*{Acknowledgements}
The authors gratefully acknowledge useful discussions with Francesco S. Cataliotti.
S.G., N.D.P., and F.C.\,were financially supported from the Fondazione CR Firenze through the project Q-BIOSCAN, PATHOS EU H2020 FET-OPEN grant no.\,828946, and UNIFI grant Q-CODYCES. M.M. acknowledges funding from the EC H2020 grant no. 820394 (ASTERIQS).


\begin{thebibliography}{10}

\bibitem{Degen2017} C. L. Degen, F. Reinhard, and P. Cappellaro, Rev. Mod. Phys. {\bf 89}, 035002 (2017).

\bibitem{Szankowski2017}
P. Sza\'{n}kowski, G. Ramon, J. Krzywda, D. Kwiatkowski, \L{}. Cywi\'{n}ski,
J. Phys.: Condensed Matter 29 (33), 333001 (2017).

\bibitem{Viola_PRL_1999}
L. Viola, E. Knill, S. Lloyd,
Phys. Rev. Lett. {\bf 82}, 2417 (1999).


\bibitem{KhodjastehPRL2005}
K. Khodjasteh, and D.A. Lidar. Fault-tolerant quantum dynamical decoupling, {\em Phys. Rev. Lett.} {\bf 95}, 180501 (2005).

\bibitem{UhrigPRL2007}
G.S. Uhrig. Keeping a Quantum Bit Alive by Optimized $\pi$-Pulse Sequences, {\em Phys. Rev. Lett.} {\bf 98}, 100504 (2007).

\bibitem{UhrigNJP2008}
G.S. Uhrig. Exact results on dynamical decoupling by $\pi$ pulses in quantum information processes, {\em New. J. Phys.} {\bf 10}, 083024 (2008).

\bibitem{Yuge2011}
T. Yuge, S. Sasaki, and Y. Hirayama,
Phys. Rev. Lett. {\bf 107}, 170504 (2011).

\bibitem{Biercuk2011}
M.J. Biercuk, A.C. Doherty, and H. Uys,
J. Phys. B: At. Mol. Opt. Phys. {\bf 44}, 154002 (2011).

\bibitem{Alvarez2011} G.A. Alvarez, and D. Suter, Phys. Rev. Lett. {\bf 107}, 230501 (2011).

\bibitem{Bylander2011}
J. Bylander, S. Gustavsson, F. Yan, F. Yoshihara, K. Harrabi, G. Fitch, D.G. Cory, Y. Nakamura, J.-S. Tsai, and W.D. Oliver,
Nat. Phys. {\bf 7}, 565-570 (2011).


\bibitem{KofmanPRL2001}
A.G. Kofman, and G. Kurizki,
Phys. Rev. Lett. {\bf 87}, 270405 (2001).

\bibitem{GordonJPB2007}
G. Gordon, N. Erez, and G. Kurizki,
J. Phys. B {\bf 40}, 75 (2007).

\bibitem{Paz-Silva2014}
G.A. Paz-Silva, and L. Viola,
Phys. Rev. Lett. {\bf 113}, 250501 (2014).

\bibitem{Zwick2016}
A. Zwick, G.A. Alvarez, and G. Kurizki,
Phys. Rev. Applied {\bf 5}, 014007 (2016).

\bibitem{Mueller2018} M.M. M\"uller, S. Gherardini, and F. Caruso, Sci. Rep. {\bf 8}, 14278 (2018).
\bibitem{Szankowski2018}
P. Sza\'{n}kowski and \L{}. Cywi\'{n}ski, Phys. Rev. A 97, 032101 (2018).

\bibitem{Krzywda2019}
J. Krzywda, P. Sza\'{n}kowski, and \L{}. Cywi\'{n}ski, New. J. Phys. 21, 043034 (2019).


\bibitem{Norris2016}
L.M. Norris, G.A. Paz-Silva, and L. Viola,
Phys. Rev. Lett. {\bf 116}, 150503 (2016).

\bibitem{Ramon2019}G. Ramon, Phys. Rev. B 100, 161302 (2019).

\bibitem{Sung2019}Y. Sung, F. Beaudoin, L. M. Norris, F. Yan, D. K. Kim, J. Y. Qiu, U. von Lüepke, J. L. Yoder, T. P. Orlando, L. Viola, et al., arXiv:1903.01043 (2019).


\bibitem{Mueller2016b} M.M. M\"uller, S. Gherardini, and F. Caruso, Sci. Rep. {\bf 6}, 38650 (2016).
\bibitem{Do2019} H.-V. Do, C. Lovecchio, I. Mastroserio, N. Fabbri, F.S. Cataliotti, S. Gherardini, M.M. M\"uller, N. Dalla Pozza, and F. Caruso, New J. Phys. {\bf 21}, 113056 (2019).

\bibitem{Sakuldee2019}
F. Sakuldee, and \L{}. Cywi{\'{n}}ski, Eprint arXiv:1907.05165 (2019).


\bibitem{Misra1977}
B. Misra, and E.C.G. Sudarshan,
J. Math. Phys. {\bf 18}, 756 (1977).

\bibitem{ItanoPRA1990}
W.M. Itano, D.J. Heinzen, J.J. Bollinger, and D.J. Wineland,
Phys. Rev. A {\bf 41}, 2295 (1990).

\bibitem{KofmanNAT2000}
A.G. Kofman, and G. Kurizki,
Nature {\bf 405}, 546--550 (2000).


\bibitem{FisherPRL2001}
M.C. Fischer, B. Gutierrez-Medina, and M.G. Raizen,
Phys. Rev. Lett. {\bf 87}, 040402 (2001).

\bibitem{FacchiPRL2002}
P. Facchi, and S. Pascazio,
Phys. Rev. Lett. {\bf 89}, 080401 (2002).

\bibitem{Facchi2008} P. Facchi, and S. Pascazio, J. Phys. A {\bf 41}, 493001 (2008).

\bibitem{SmerziPRL2012}
A. Smerzi,
Phys. Rev. Lett. {\bf 109}, 150410 (2012).

\bibitem{SchaferZeno}
F. Sch\"afer, I. Herrera, S. Cherukattil, C. Lovecchio, F.S. Cataliotti, F. Caruso, A. Smerzi,
Nat. Commun. {\bf 5}, 3194 (2014).

\bibitem{Signoles2014}
A. Signoles, A. Facon, D. Grosso, I. Dotsenko, S. Haroche, J. Raimond, M. Brune, and S. Gleyzes,
Nat. Phys. {\bf 10}, 715--719 (2014).


\bibitem{Goan}
Z. Zhou, Z. L\"{u}, H. Zheng, and H.-S. Goan,  {\em Phys. Rev. A} {\bf 96}, 032101 (2017).

\bibitem{Hanggi}
L. Magazz\`{u}, P. Talkner and Peter H\"{a}nggi, {\em New J. Phys.} \textbf{20}, 033001 (2018).

\bibitem{Chaudhry}
A. Z. Chaudhry, {\em Sci. Rep.}{\bf 6}, 29497 (2016).



\bibitem{Gherardini2016} S. Gherardini, S. Gupta, F.S. Cataliotti, A. Smerzi, F. Caruso, and S. Ruffo, New J. Phys. {\bf 18}, 013048 (2016).
\bibitem{Mueller2017} M.M. M\"uller, S. Gherardini, and F. Caruso, Annalen der Physik {\bf 529 (9)}, 1600206 (2017).

\bibitem{Gherardini2017} S. Gherardini, C. Lovecchio, M.M. M\"uller, P. Lombardi, F. Caruso, and F.S. Cataliotti, Quantum Science and Technology {\bf 2 (1)}, 015007 (2017).
\bibitem{Mueller2016} M.M. M\"uller, S. Gherardini, A. Smerzi, and F. Caruso, Phys. Rev. A {\bf 94}, 042322 (2016).


\bibitem{Balasubramanian2019}
P. Balasubramanian, C. Osterkamp, Y. Chen, X. Chen, T. Teraji, E. Wu, B. Naydenov, and F. Jelezko,
Nano Letters 19 (9), 6681-6686 (2019).

\bibitem{Zhou2019}
H. Zhou, J. Choi, S. Choi, R. Landig, A. M. Douglas, J. Isoya, F. Jelezko, S. Onoda, H. Sumiya, P. Cappellaro, H. S. Knowles, H. Park, M. D. Lukin,
arXiv:1907.10066 (2019).



\bibitem{DallaPozza2019}N. Dalla Pozza, M.M. M\"uller, S. Gherardini, and F. Caruso, arXiv:1911.10598 (2019).
\bibitem{Ketterle}
W. Ketterle, D.S. Durfee and D.M. Stamper-Kurn, \textit{Making, probing and understanding Bose-Einstein condensates} in Proceedings of the International School of Physics "Enrico Fermi" Volume 140, p. 67, (1999).





\end{thebibliography}

\end{document}